\documentclass[11pt,onecolumn]{article}

\usepackage{bm}
\usepackage{amsfonts}
\usepackage{amsmath}
\usepackage{amssymb}
\usepackage{cite}
\usepackage[amsmath,thmmarks]{ntheorem}

\usepackage{color, soul}
\usepackage{epic, eepic}
\usepackage[top=1in, bottom=1in, left=1.25in, right=1.25in]{geometry}
\usepackage{appendix}

  \usepackage[pdftex]{graphicx}
  \graphicspath{{../pdf/}{../jpeg/}}
  \DeclareGraphicsExtensions{.pdf,.jpeg,.png}

\usepackage{color, soul}
\usepackage{array}
\usepackage{theorem}
\usepackage{booktabs}

\newtheorem{definition}{Definition}
\newtheorem{lemma}{Lemma}
\newtheorem{theorem}{Theorem}

\theoremheaderfont{\sc}\theorembodyfont{\upshape}
\theoremstyle{nonumberplain}
\theoremseparator{}
\theoremsymbol{\rule{1ex}{1ex}}
\newtheorem{proof}{Proof}

\begin{document}

\title{Oracle-order Recovery Performance of Greedy Pursuits with Replacement against General Perturbations}

\author{Laming~Chen
        and~Yuantao~Gu\thanks{This work was supported partially by the National Natural
 Science Foundation of CHINA (60872087 and U0835003) and Agilent
 Technologies Foundation \# 2205. The authors are with Department
 of Electronic Engineering, Tsinghua University, Beijing 100084,
 China. The corresponding author of this paper is Yuantao Gu (Email: gyt@tsinghua.edu.cn).}}

\date{Received Dec. 30, 2011; revised Jan. 20, 2013.}

\maketitle

\begin{abstract}
Applying the theory of compressive sensing in practice always takes different kinds of perturbations into consideration. In this paper, the recovery performance of greedy pursuits with replacement for sparse recovery is analyzed when both the measurement vector and the sensing matrix are contaminated with additive perturbations. Specifically, greedy pursuits with replacement include three algorithms, compressive sampling matching pursuit (CoSaMP), subspace pursuit (SP), and iterative hard thresholding (IHT), where the support estimation is evaluated and updated in each iteration. Based on restricted isometry property, a unified form of the error bounds of these recovery algorithms is derived under general perturbations for compressible signals. The results reveal that the recovery performance is stable against both perturbations. In addition, these bounds are compared with that of oracle recovery--- least squares solution with the locations of some largest entries in magnitude known a priori. The comparison shows that the error bounds of these algorithms only differ in coefficients from the lower bound of oracle recovery for some certain signal and perturbations, as reveals that oracle-order recovery performance of greedy pursuits with replacement is guaranteed. Numerical simulations are performed to verify the conclusions.

\textbf{Keywords:} Compressive sensing, \, sparse recovery, \, general perturbations, \, performance analysis, \, restricted isometry property, \, greedy pursuits, \, compressive sampling matching pursuit, \, subspace pursuit, \, iterative hard thresholding, \, oracle recovery.
\end{abstract}

\section{Introduction}

Compressive sensing, or compressive sampling (CS) \cite{Candes,Donoho,Tao}, is a novel signal processing technique proposed to effectively sample and compress sparse signals, i.e., signals that can be represented by few significant coefficients in some basis. Assume that the signal of interest ${\bf x}\in\mathbb{R}^N$ can be represented by $\bf x=\Psi s$, where ${\bf \Psi}\in\mathbb{R}^{N\times N}$ is the basis matrix and ${\bf s}$ is $K$-sparse, which means only $K$ out of its $N$ entries are nonzero. One of the essential issues of CS theory lies in recovering ${\bf x}$ (or equivalently, $\bf s$) from its linear observations,
\begin{align}\label{y=Ax}
{\bf y}={\bf \Phi x}={\bf \Phi \Psi s},
\end{align}
where ${\bf \Phi}\in\mathbb{R}^{M\times N}$ is a sensing matrix with more columns than rows and ${\bf y}$ is the measurement vector. Unfortunately, directly finding the sparsest solution to (\ref{y=Ax}) is NP-hard, which is not practical for sparse recovery. This leads to one of the major aspects of CS theory---designing effective recovery algorithms with low computational complexity and fine recovery performance.

\subsection{Overview of Sparse Recovery Algorithms}

A family of convex relaxation algorithms for sparse recovery \cite{BP} had been introduced before the theory of CS was established. Based on linear programming (LP) techniques, it is shown that $\ell_1$ norm optimization, also known as basis pursuit (BP),
\begin{align}\label{BP}
\min_{\bf s} \left\|{\bf s}\right\|_1\quad{\rm s.t.}\quad{\bf y}={\bf \Phi \Psi}{\bf s}
\end{align}
yields the sparse solution as long as $\bf A=\Phi \Psi$ satisfies the restricted isometry property (RIP) with a constant parameter \cite{LP,RIP,newbounds}. Recovery algorithms based on convex optimization include interior-point methods \cite{convexop,l1ls} and homotopy methods \cite{Osborne,Efron}.

In contrast with convex relaxation algorithms, non-convex optimization algorithms solve (\ref{y=Ax}) by minimizing $\ell_p$ norm with respect to $0<p<1$, which is not convex. Typical algorithms include focal underdetermined system solver (FOCUSS) \cite{FOCUSS}, iteratively reweighted least squares (IRLS) \cite{IRLS}, smoothed $\ell^0$ (SL0) \cite{SL0}, and zero-point attracting projection (ZAP) \cite{ZAP,l1ZAP}. Compared with convex relaxation algorithms, theoretical analysis based on RIP shows that fewer measurements are required for exact recovery by non-convex optimization methods \cite{nonconvex}.

A family of iterative greedy algorithms has received much attention due to their simple implementation and low computational complexity. The basic idea underlying these algorithms is to iteratively estimate the support set of the unknown sparse signal, i.e., the set of locations of its nonzero entries. In each iteration, one or more indices are added to the support estimation by correlating the columns of $\bf A$ with the regularized measurement vector. Typical examples include orthogonal matching pursuit (OMP) \cite{OMP}, regularized OMP (ROMP) \cite{ROMP}, and stage-wise OMP (StOMP) \cite{StOMP}. Compared with convex relaxation algorithms, greedy pursuits need more measurements, but they tend to be more computationally efficient.

Recently, several greedy algorithms including compressive sampling matching pursuit (CoSaMP) \cite{CoSaMP} and subspace pursuit (SP) \cite{SP} have been proposed by incorporating the idea of backtracking. In each iteration, SP algorithm refines the $K$ columns of matrix $\bf A$ that span the subspace where the measurement vector $\bf y$ lies. Specifically, SP adds $K$ more indices to the $K$ candidates of support estimate, and discards the most unreliable $K$ ones. Similarly, CoSaMP adds $2K$ more indices in each iteration, while computes the regularized measurement vector in a different way. By evaluating the reliability of all candidates in each iteration, these algorithms can provide comparable performance to convex relaxation algorithms, and exhibit low computational complexity as matching pursuit algorithms.

Another kind of greedy pursuits, including iterative hard thresholding (IHT) \cite{IHT} and its normalized variation \cite{NIHT}, is proposed with the advantages of low computational complexity and theoretical performance guarantee. In each iteration, the entries of the iterative solution except for the most $K$ reliable ones are set to zero. Together with CoSaMP and SP, these algorithms can be considered as greedy pursuits with replacement involved in each iteration.

\subsection{Overview of Perturbation Analysis}

To apply the theory of CS in practice, the effect of noise and perturbation must be taken into consideration. The common analysis is the additive noise to the measurements, i.e.,
\begin{align}\label{meapur}
{\bf \tilde y}={\bf y}+{\bf e},
\end{align}
where $\bf e$ is termed measurement noise. Most existing algorithms have been proved to be stable in this scenario, including theoretical analysis of BP \cite{RIP}, ROMP \cite{ROMPnoise}, CoSaMP \cite{CoSaMP}, SP \cite{SP}, and IHT \cite{IHT,NIHT}. It is shown that the error bounds of the recovered solutions are proportional to the $\ell_2$ norm of $\bf e$. A certain distribution of the measurement noise can be introduced to achieve better results, such as Gaussian \cite{Dantzig,Bickel,Ben-Haim,Giryes} or others \cite{Rish,Possion}.

Until recently, only a few researches involve the perturbation to the sensing matrix $\bf \Phi$, which is also termed system perturbation. Existing works include analysis of BP \cite{Matthew}, CoSaMP \cite{Herman}, SP \cite{Wang}, $\ell_p$ norm minimization with $p\in(0,1]$ \cite{Aldroubi}, and OMP \cite{Dingjie}. In these works, (\ref{meapur}) is extended by introducing a perturbed sensing matrix, i.e., ${\bf \tilde \Phi}={\bf \Phi}+{\bf \Delta}$. It is of great significance to analyze the stability of recovery algorithms against both perturbations when the theory of CS is applied in practice. Other related works include mismatch of sparsity basis \cite{Mismatch} and sparsity-cognizant total least-squares \cite{Zhu}.

Two practical scenarios are usually considered in CS applications. First, when $\bf \Phi$ represents a system model, $\bf \Delta$ denotes the precision error when the system is physically implemented. Thus the whole sensing process is
\begin{align}\label{sysper1}
{\bf \tilde y}=\left({\bf \Phi}+{\bf \Delta}\right){\bf x}+{\bf e},
\end{align}
and only the nominal sensing matrix and contaminated measurement vector are available for recovery, i.e., ${\bf \hat x}={\rm R}({\bf \Phi},{\bf \tilde y})$ where ${\rm R}(\cdot)$ denotes a certain recovery algorithm.

In countless other problems, $\bf \Delta$ is involved due to mis-modeling of the actual system $\bf \Phi$. Thus ${\bf \tilde \Phi}={\bf \Phi}+{\bf \Delta}$ and the sensing process is
\begin{align}\label{sysper2}
{\bf \tilde y}={\bf \Phi}{\bf x}+{\bf e}.
\end{align}
Both the sensing matrix and measurement vector are contaminated, and the recovered solution is ${\bf \hat x}={\rm R}({\bf \tilde \Phi},{\bf \tilde y})$.

Existing works \cite{Matthew,Herman,Wang,Dingjie} are all based on the latter scenario, thus it is well considered and fully analyzed in this paper. The first scenario (\ref{sysper1}) is briefly discussed after that as a remark.

\subsection{Our Work}

This paper mainly considers the recovery performance of greedy pursuits with replacement against general perturbations. Specifically, when both measurement noise and system perturbation exist, the error bounds of the solutions of CoSaMP, SP, and IHT are derived and discussed in detail for not strictly sparse signals, i.e., compressible signals. It is shown that these relative error bounds are linear in the relative perturbations of the measurement vector and the sensing matrix, which indicates that the recovery performance is stable against general perturbations. These error bounds are also compared with that of oracle recovery, and it reveals that the results in this paper are optimal up to the coefficients. Numerical simulations are performed to verify the conclusions in this paper. Previous related works including \cite{Giryes,Herman,Wang} are compared with this work in Section VI.

The remainder of this paper is organized as follows. Section II gives a brief review of RIP and three greedy pursuits with replacement. Section III presents the main theorems about the recovery performance of greedy pursuits with replacement against general perturbations, and the results are shown to be of the same order as oracle recovery by comparing them in Section IV. Numerical simulations are performed in Section V. Several related works are discussed in Section VI, and the paper is concluded in Section VII. Some detailed descriptions and discussions of greedy pursuits with replacement and proofs are postponed to Appendix.

\section{Preliminary}

In this section, the definition of RIP and descriptions of several greedy pursuits, including CoSaMP, SP, and IHT, as well as their recovery performance against measurement noise, are introduced.

\subsection{Restricted Isometry Property}

The restricted isometry property (RIP) for any matrix $\bf A$ describes the degree of orthogonality among its different columns.

\begin{definition}
\cite{LP} For positive integer $K$, define the restricted isometry constant (RIC) $\delta_K$ of a matrix $\bf A$ as the smallest non-negative number such that
\begin{align}\label{RIP}
\left(1-\delta_K\right)\left\|{\bf s}\right\|_2^2\leq\left\|{\bf A}{\bf s}\right\|_2^2 \leq\left(1+\delta_K\right)\left\|{\bf s}\right\|_2^2
\end{align}
holds for any $K$-sparse vector $\bf s$.
\end{definition}

According to Definition~1, $\delta_K<1$ implies that every $K$ columns of $\bf A$ are linearly independent, and $\delta_{2K}\ll 1$ implies that $\bf A$ almost maintains the $\ell_2$ distance between any pair of $K$-sparse signals. It is also easy to check that if $\bf A$ satisfies the RIP with $\delta_{K_1}$ and $\delta_{K_2}$, and $K_1\leq K_2$, then $\delta_{K_1}\leq\delta_{K_2}$.

Calculating the exact value of RIC is intractable because it involves all submatrices comprised of $K$ columns of $\bf A$. Fortunately, random matrices possess small RICs for overwhelming probability. For example, if the entries of ${\bf A}\in \mathbb{R}^{M\times N}$ are independently and identically distributed Gaussian random variables with zero mean and variance $1/M$, and $M$ satisfies
\begin{align*}
M\geq \frac{\displaystyle CK\log{(N/K)}}{\displaystyle \varepsilon^2},
\end{align*}
then $\delta_K\leq \varepsilon$ with at least probability $1-{\rm e}^{-cM}$, where $C$ and $c$ are two constants \cite{LP,CoSaMP}. Recent theoretical results about the bounds of RIC can be found in \cite{Bah,Blanchard}.

Though RIP is usually used to show the impact of matrix on sparse signals, the following lemma permits us to generalize the result from sparse signals to general signals.

\begin{lemma}\label{RIPgeneral}
(Proposition~3.5 in \cite{CoSaMP}) Suppose that the matrix $\bf A$ satisfies RIP of level $K$ with $\delta_{K}$, then for any signal ${\bf s}\in\mathbb{R}^N$, it holds that
\begin{align}
\left\|{\bf As}\right\|_2\leq\sqrt{1+\delta_K}\left(\left\|{\bf s}\right\|_2+\frac{\displaystyle 1}{\displaystyle \sqrt{K}}\left\|{\bf s}\right\|_1\right).
\end{align}
\end{lemma}

Lemma~\ref{RIPgeneral} is used in the theoretical analysis of this paper for compressible signals.

\subsection{Greedy Pursuits with Replacement}

Greedy pursuits with replacement include CoSaMP, SP, and IHT algorithms, and they are briefly introduced as follows. The pseudo codes for these algorithms and some discussions are postponed to Appendix~A.

The CoSaMP algorithm iteratively refines the support of $K$-sparse vector $\bf s$. In each iteration, $2K$ more indices are selected by correlating $\bf A$ with the residue vector $\bf r$, then the best $K$ candidates out of at most $3K$ ones are kept and the residue vector is updated. The details of CoSaMP can be found in \cite{CoSaMP}.

The SP algorithm is firstly proposed in \cite{Dai} and further developed in \cite{SP}. Unlike CoSaMP, SP adds $K$ more indices in each iteration, and keeps the best $K$ candidates. In addition, the residue vector of SP is orthogonal to the subspace spanned by the columns of $\bf A$ indexed by the $K$ candidates, making the new $K$ indices added in each iteration totally different from the previously identified $K$ ones. Please refer to \cite{SP} for more details.

The IHT algorithm is firstly proposed in \cite{IHTfirst}, and later developed and analyzed in \cite{IHT}. To improve the convergence performance of the method, a normalized variation NIHT is proposed in \cite{NIHT} which retains theoretical performance guarantee similar to IHT. Without loss of generality, only IHT algorithm is discussed in the following analysis.

The following Theorem~\ref{GPRa} reveals the error bounds of the solutions of greedy pursuits with replacement when only measurement noise exists. For two sets ${\rm S}_1$ and ${\rm S}_2$, ${\rm S}_1-{\rm S}_2$ denotes the set comprised of all elements $x\in{\rm S}_1$ and $x\notin{\rm S}_2$. ${\bf s}_{\rm S}$ denotes the subvector composed of entries of $\bf s$ indexed by set ${\rm S}$.

\begin{theorem}\label{GPRa}
Given a noisy measurement vector ${\bf \tilde y}={\bf As}+{\bf e}$ where ${\bf s}$ is $K$-sparse, the estimated solution ${\bf s}^{[l]}$ in the $l$-th iteration of CoSaMP and IHT algorithms satisfies
\begin{align}\label{GPRerror1}
\big\|{\bf s}-{\bf s}^{[l]}\big\|_2\leq C\big\|{\bf s}-{\bf s}^{[l-1]}\big\|_2 +C_1\left\|{\bf e}\right\|_2,
\end{align}
and the estimated support set ${\rm S}^l$ in the $l$-th iteration of SP algorithm satisfies
\begin{align}\label{SPerror1}
\left\|{\bf s}_{{\rm S}-{\rm S}^l}\right\|_2\leq C\left\|{\bf s}_{{\rm S}-{\rm S}^{l-1}}\right\|_2+C_1\left\|{\bf e}\right\|_2,
\end{align}
where $\rm S$ denotes the support of $\bf s$.

Furthermore, if the matrix $\bf A$ satisfies $\delta_{bK}\leq c$, then $C<1$, and it can be derived that
\begin{align}\label{GPRerror2}
\big\|{\bf s}-{\bf s}^{[l]}\big\|_2\le aC^l\left\|{\bf s}\right\|_2 +D\left\|{\bf e}\right\|_2
\end{align}
holds for greedy pursuits with replacement. The specific values of the constants $a$, $b$, $c$, $C$, $C_1$, and $D$ are illustrated in TABLE~\ref{tableconstant1}.
\end{theorem}

\begin{table}[t]
\renewcommand{\arraystretch}{1.8}
\caption{The Specification of the Constants}
\begin{center}\label{tableconstant1}
\begin{tabular}{cccc}
\toprule[1pt]
 & CoSaMP & SP & IHT\\
\hline
$a$ & 1 & 1.26 & 1\\
$b$ & 4 & 3 & 3\\
$c$ & 0.171& 0.206 & 0.353\\
$C$ & $\frac{\displaystyle 4\delta_{4K}}{\displaystyle (1-\delta_{4K})^2}$ & $\frac{\displaystyle 2\delta_{3K}+2\delta_{3K}^2}{\displaystyle (1-\delta_{3K})^3}$ & $\sqrt{8}\delta_{3K}$\\
$C_1$ & $\frac{\displaystyle 6+2\delta_{4K}}{\displaystyle (1-\delta_{4K})^2}$ & $\frac{\displaystyle 4(1+\delta_{3K})}{\displaystyle (1-\delta_{3K})^2}$ & $2\sqrt{1+\delta_{3K}}$\\
$D$ & $\frac{\textstyle6+2\delta_{4K}}{\textstyle1-6\delta_{4K}+\delta_{4K}^2}$ & $\frac{\textstyle5-4\delta_{3K}-8\delta_{3K}^2-\delta_{3K}^4}{\textstyle1-6\delta_{3K}+6\delta_{3K}^2-2\delta_{3K}^3+\delta_{3K}^4}$ & $\frac{\textstyle2\sqrt{1+\delta_{3K}}}{\textstyle1-\sqrt{8}\delta_{3K}}$\\
\bottomrule[1pt]
\end{tabular}
\end{center}
\end{table}

\begin{proof}
Theorem~\ref{GPRa} is concluded from the results of \cite{CoSaMP,SP,IHT}. The proof is postponed to Appendix B.
\end{proof}

Now we have introduced CoSaMP, SP, and IHT algorithms, where sparsity level $K$ needs to be known a priori for replacement. According to (\ref{GPRerror2}), when only measurement noise exists, the error bounds of the solutions are exponential decay function of iteration number, and the limits of them are proportional to the $\ell_2$ norm of the noise. Some remarks based on (\ref{GPRerror2}) are derived in Appendix~A, and the main conclusions in this paper are based on (\ref{GPRerror2}) as well. It needs to be emphasized that the analysis is performed as worst-case, and the demands of RICs of $\bf A$ are sufficient conditions. Better recovery performance of these three algorithms is normally achieved in practice, thus they are widely applied in various applications.

\section{Recovery Performance of Greedy Pursuits with Replacement}

In this section, the error bounds of greedy pursuits with replacement are derived for general signals when both measurement noise and system perturbation exist.

\subsection{Notations and Assumptions}

Several notations and assumptions are stated first and they will be used in the following analysis. ${\bf A}_{\rm S}$ denotes the submatrix composed of the columns of $\bf A$ indexed by set $\rm S$. ${\bf A}^{\rm T}$ denotes the transpose of matrix $\bf A$. ${\bf A}^{\dagger}$ denotes the pseudo-inverse of matrix $\bf A$. The signal $\bf s$ is assumed to be a compressible signal in the following context, which means that it can be well approximated by a sparse signal. Vector ${\bf s}_K$ is assumed to be the best $K$-term approximation to $\bf s$, and define the approximation error ${\bf s}_K^c={\bf s}-{\bf s}_K$. The approximation error can be quantified as
\begin{align}\label{rs}
r_{K}=\frac{\displaystyle \left\|{\bf s}_K^c\right\|_2}{\displaystyle \left\|{\bf s}\right\|_2},\quad
s_{K}=\frac{\displaystyle \left\|{\bf s}_K^c\right\|_1}{\displaystyle \sqrt{K}\left\|{\bf s}\right\|_2}.
\end{align}
When $\bf s$ is $K$-sparse, the ratios $r_{K}$ and $s_{K}$ are both zero. If $\bf s$ is compressible, then for reasonable $K$, the ratios are expected to be far less than 1.

The symbol $\|{\bf A}\|_2$ denotes the spectral norm of a matrix $\bf A$. The measurement noise $\bf e$ and system perturbation $\bf \Delta$ can be quantified with the following relative perturbations
\begin{align}\label{errorboundsss}
\varepsilon_{\bf y}=\frac{\displaystyle \left\|{\bf e}\right\|_2}{\displaystyle \left\|{\bf y}\right\|_2},\quad
\varepsilon_{\bf \Phi}=\frac{\displaystyle \left\|{\bf \Delta}\right\|_2}{\displaystyle \left\|{\bf \Phi}\right\|_2},
\end{align}
where $\|{\bf y}\|_2$ and $\|{\bf \Phi}\|_2$ are nonzero. $\varepsilon_{\bf y}$ and $\varepsilon_{\bf \Phi}$ quantify the relative measurement noise and relative system perturbation, respectively. In practical scenarios, the exact forms of $\bf e$ and $\bf \Delta$ are not known in advance, thus the relative perturbations are applied instead. In this paper, $\varepsilon_{\bf y}$ and $\varepsilon_{\bf \Phi}$ are assumed less than 1. Define
\begin{align}
\bf \tilde A=(\Phi+\Delta)\Psi=\Phi\Psi+\Delta\Psi\triangleq A+E.
\end{align}
Let $\|{\bf A}\|_2^{(K)}$ denote the largest spectral norm taken over all $K$-column submatrices of $\bf A$. Assume
\begin{align}\label{matrixboundsss}
\varepsilon_{\bf A}=\frac{\displaystyle \left\|{\bf E}\right\|_2}{\displaystyle \left\|{\bf A}\right\|_2},\quad
\varepsilon_{\bf A}^{(K)}=\frac{\displaystyle \left\|{\bf E}\right\|_2^{(K)}}{\displaystyle \left\|{\bf A}\right\|_2^{(K)}},
\end{align}
which also represent the relative system perturbation. It is easy to derive that
\begin{align*}
\varepsilon_{\bf A}\le\kappa_{\bf \Psi}\varepsilon_{\bf \Phi},\quad
\varepsilon_{\bf A}^{(K)}\le\frac{\left\|{\bf \Phi}\right\|_2\left\|{\bf \Psi}\right\|_2}{\sqrt{1-\delta_K}}\varepsilon_{\bf A},
\end{align*}
where $\kappa_{\bf \Psi}=\left\|{\bf \Psi}\right\|_2\left\|{\bf \Psi}^{-1}\right\|_2$ is the condition number of $\bf \Psi$. Particularly, $\kappa_{\bf \Psi}$ is equal to one if $\bf x$ is sparse in orthogonal basis. In general cases, the relative perturbations (\ref{matrixboundsss}) are approximately the same as $\varepsilon_{\bf \Phi}$ if $\kappa_{\bf \Psi}$ is not very large. The following lemma quantifies the RICs of the matrix ${\bf \tilde A}$.

\begin{lemma}\label{RIPerror}
(Theorem~1 in \cite{Matthew}) Assume the matrix $\bf A$ satisfies RIP of level $K$ with $\delta_K$, and the relative perturbation $\varepsilon_{\bf A}^{(K)}$ is associated with matrix $\bf E$. Define the constant
\begin{align}
{\tilde \delta}_{K,\max}\triangleq \left(1+\delta_K\right)\left(1+\varepsilon_{\bf A}^{(K)}\right)^2-1,
\end{align}
then the RIC ${\tilde\delta}_K$ of matrix ${\bf \tilde A}={\bf A}+{\bf E}$ satisfies ${\tilde\delta}_K\leq{\tilde \delta}_{K,\max}$.
\end{lemma}

As can be seen from Lemma~\ref{RIPerror}, the upper bound of the RIC of perturbed matrix ${\bf \tilde A}$ is slightly bigger than that of $\bf A$ if the relative perturbation $\varepsilon_{\bf A}^{(K)}$ is small. Notice that ${\tilde \delta}_{K,\max}$ represents a worst-case of ${\tilde \delta}_{K}$, thus better RICs of ${\bf \tilde A}$ are normally achieved in practice.

\subsection{Error Bounds of Greedy Pursuits with Replacement}

The relative error bounds of greedy pursuits with replacement under general perturbations for compressible signals are given in this subsection. The following Theorem~\ref{Greedyerrorth} summarizes the main result.

\begin{theorem}\label{Greedyerrorth}
Suppose ${\bf y}={\bf \Phi}{\bf x}$, where ${\bf x}={\bf \Psi s}$ and $\bf s$ is a compressible vector. The available information for recovering $\bf x$ is ${\bf \tilde y}={\bf y}+{\bf e}$, ${\bf \tilde \Phi}={\bf \Phi}+{\bf \Delta}$, and basis matrix $\bf \Psi$. If the available perturbed matrix ${\bf \tilde A}={\bf \tilde \Phi}{\bf \Psi}$ satisfies RIP with
\begin{align}\label{generalcondition}
{\tilde\delta}_{bK}\le c,
\end{align}
and the non-perturbed matrix $\bf A$ satisfies RIP of level $K$ with $\delta_K$, then in the $l$-th iteration, the relative error of the solution ${\bf x}^{[l]}={\bf \Psi}{\bf s}^{[l]}$ of greedy pursuits with replacement satisfies
\begin{align}\label{generalrelativeerror}
\frac{\displaystyle \big\|{\bf x}-{\bf x}^{[l]}\big\|_2}{\displaystyle \left\|{\bf x}\right\|_2}&\le\kappa_{\bf \Psi}\bigg(a{\tilde C}^{l}+r_{K}+{\tilde D}\sqrt{1+\delta_K}\left(\varepsilon_{\bf y}+\varepsilon_{\bf A}^{(K)}
+\left(1+\varepsilon_{\bf y}\right)\left(r_{K}+s_{K}\right)\right)\bigg),
\end{align}
where the specific values of the constants $a$, $b$, $c$, $\tilde C$, and $\tilde D$ are illustrated in TABLE~\ref{tableconstant1} and TABLE~\ref{tableconstant}.

Furthermore, after at most
\begin{align}\label{generaliteration}
l=\left\lceil\log_{\tilde C}\left(\frac{\displaystyle \varepsilon_{\bf y}+\varepsilon_{\bf A}^{(K)}+s_K}{\displaystyle a}\right)\right\rceil
\end{align}
iterations, these algorithms estimate $\bf x$ with accuracy
\begin{align}\label{generalrelativeerroriteration}
\frac{\displaystyle \big\|{\bf x}-{\bf x}^{[l]}\big\|_2}{\displaystyle \left\|{\bf x}\right\|_2}\le &\kappa_{\bf \Psi}\left({\tilde D}\sqrt{1+\delta_K}+1\right)\left(\varepsilon_{\bf y}+\varepsilon_{\bf A}^{(K)}+\left(1+\varepsilon_{\bf y}\right)\left(r_{K}+s_{K}\right)\right).
\end{align}
\end{theorem}

\begin{table}[t]
\renewcommand{\arraystretch}{2}
\caption{The Specification of the Constants}
\begin{center}\label{tableconstant}
\begin{tabular}{cccc}
\toprule[1pt]
 & CoSaMP & SP & IHT\\
\hline
${\tilde C}$ & $\frac{\displaystyle 4{\tilde\delta}_{4K}}{\displaystyle (1-{\tilde\delta}_{4K})^2}$ & $\frac{\displaystyle 2{\tilde\delta}_{3K}+2{\tilde\delta}_{3K}^2}{\displaystyle (1-{\tilde\delta}_{3K})^3}$ & $\sqrt{8}{\tilde\delta}_{3K}$\\
${\tilde D}$ & $\frac{\textstyle6+2{\tilde\delta}_{4K}}{\textstyle1-6{\tilde\delta}_{4K}+{\tilde\delta}_{4K}^2}$ & $\frac{\textstyle 5-4{\tilde\delta}_{3K}-8{\tilde\delta}_{3K}^2-{\tilde\delta}_{3K}^4} {\textstyle1-6{\tilde\delta}_{3K}+6{\tilde\delta}_{3K}^2-2{\tilde\delta}_{3K}^3+{\tilde\delta}_{3K}^4}$ & $\frac{\textstyle2\sqrt{1+{\tilde\delta}_{3K}}}{\textstyle1-\sqrt{8}{\tilde\delta}_{3K}}$\\
\bottomrule[1pt]
\end{tabular}
\end{center}
\end{table}

For exact sparse signals, a better and more intuitive result can be derived from Theorem~\ref{Greedyerrorth} by setting $r_K$ and $s_K$ to zero. The result is stated as Theorem~\ref{greedyerrorsparseth}.

\begin{theorem}\label{greedyerrorsparseth}
Suppose ${\bf y}={\bf \Phi}{\bf x}$, where ${\bf x}={\bf \Psi s}$ and $\bf s$ is $K$-sparse. The available information for recovering $\bf x$ is ${\bf \tilde y}={\bf y}+{\bf e}$, ${\bf \tilde \Phi}={\bf \Phi}+{\bf \Delta}$, and basis matrix $\bf \Psi$. If the available perturbed matrix ${\bf \tilde A}={\bf \tilde \Phi}{\bf \Psi}$ satisfies RIP with
\begin{align}
{\tilde\delta}_{bK}\le c,
\end{align}
and the non-perturbed matrix $\bf A$ satisfies RIP of level $K$ with $\delta_K$, then in the $l$-th iteration, the relative error of the solution ${\bf x}^{[l]}={\bf \Psi}{\bf s}^{[l]}$ of greedy pursuits with replacement satisfies
\begin{align}\label{sparserelativeerror}
\hspace{-0.05in}\frac{\displaystyle \big\|{\bf x}-{\bf x}^{[l]}\big\|_2}{\displaystyle \left\|{\bf x}\right\|_2}\le \kappa_{\bf \Psi}\left(a{\tilde C}^{l}+{\tilde D}\sqrt{1+\delta_K}\left(\varepsilon_{\bf y}+\varepsilon_{\bf A}^{(K)}\right)\right),
\end{align}
where the specific values of the constants $a$, $b$, $c$, $\tilde C$, and $\tilde D$ are illustrated in TABLE~\ref{tableconstant1} and TABLE~\ref{tableconstant}.

Furthermore, after at most
\begin{align}
l=\left\lceil\log_{\tilde C}\left(\frac{\displaystyle \varepsilon_{\bf y}+\varepsilon_{\bf A}^{(K)}}{\displaystyle a}\right)\right\rceil
\end{align}
iterations, these algorithms estimate $\bf x$ with accuracy
\begin{align}\label{sparserelativeerroriteration}
\frac{\displaystyle \big\|{\bf x}-{\bf x}^{[l]}\big\|_2}{\displaystyle \left\|{\bf x}\right\|_2}\le \kappa_{\bf \Psi}\left({\tilde D}\sqrt{1+\delta_K}+1\right)\left(\varepsilon_{\bf y}+\varepsilon_{\bf A}^{(K)}\right).
\end{align}
\end{theorem}

\subsection{Proof of Theorem~\ref{Greedyerrorth}}

\begin{proof}
Recalling the sensing process (\ref{sysper2}) with general perturbations, it is equivalent to
\begin{align}
{\bf \tilde y}&={\bf A}{\bf s}+{\bf e}=\left({\bf\tilde A}-{\bf E}\right){\bf s}_K+{\bf A}{\bf s}_K^c+{\bf e}\nonumber\\
&={\bf \tilde A}{\bf s}_K+\left({\bf e}-{\bf E}{\bf s}_K+{\bf A}{\bf s}_K^c\right).
\end{align}
Define ${\bf \tilde e}={\bf e}-{\bf E}{\bf s}_K+{\bf A}{\bf s}_K^c$ as the error term. According to Theorem~\ref{GPRa}, under condition (\ref{generalcondition}) about the RICs of ${\bf \tilde A}$, the solution ${\bf s}^{[l]}$ in the $l$-th iteration satisfies
\begin{align}\label{proofequ1}
\big\|{\bf s}_K-{\bf s}^{[l]}\big\|_2\le a{\tilde C}^{l}\left\|{\bf s}_K\right\|_2 +{\tilde D}\left\|{\bf\tilde e}\right\|_2,
\end{align}
where ${\tilde C}<1$ and $\tilde D$ are constants specified in TABLE~\ref{tableconstant}. From the triangle inequality, it can be derived that
\begin{align}\label{proofequ2}
\big\|{\bf s}-{\bf s}^{[l]}\big\|_2\leq\left\|{\bf s}-{\bf s}_K\right\|_2+\big\|{\bf s}_K-{\bf s}^{[l]}\big\|_2.
\end{align}
Substituting (\ref{proofequ1}) into (\ref{proofequ2}), and from the fact that
\begin{align}\label{proofequ8}
\left\|{\bf s}_K\right\|_2\leq\left\|{\bf s}\right\|_2,
\end{align}
the error bound in the $l$-th iteration satisfies
\begin{align}\label{proofequ3}
\big\|{\bf s}-{\bf s}^{[l]}\big\|_2\le a{\tilde C}^{l}\left\|{\bf s}\right\|_2+\left\|{\bf s}_K^c\right\|_2+{\tilde D}\left\|{\bf\tilde e}\right\|_2.
\end{align}

To estimate $\|{\bf \tilde e}\|_2$, the triangle inequality, Lemma~\ref{RIPgeneral}, and definitions (\ref{rs}) imply
\begin{align}\label{proofequ4}
\left\|{\bf \tilde e}\right\|_2 \leq&\left\|{\bf e}\right\|_2+\left\|{\bf E}{\bf s}_K\right\|_2+\left\|{\bf A}{\bf s}_K^c\right\|_2\nonumber\\
\leq&\left\|{\bf e}\right\|_2+\left\|{\bf E}\right\|_2^{(K)}\left\|{\bf s}_K\right\|_2+\left\|{\bf A}\right\|_2^{(K)}\left\|{\bf s}\right\|_2\left(r_K+s_K\right).
\end{align}
It can be derived that
\begin{align}\label{proofequ6}
\left\|{\bf e}\right\|_2=&\varepsilon_{\bf y}\left\|{\bf As}\right\|_2 \leq\varepsilon_{\bf y}\left(\left\|{\bf As}_K\right\|_2+\left\|{\bf A}{\bf s}_K^c\right\|_2 \right)\nonumber\\
\leq&\varepsilon_{\bf y}\left\|{\bf A}\right\|_2^{(K)}\left\|{\bf s}\right\|_2 \left(1+r_{K}+s_{K}\right),
\end{align}
which indicates
\begin{align}\label{proofequ7}
\left\|{\bf \tilde e}\right\|_2\le\left\|{\bf A}\right\|_2^{(K)}\left\|{\bf s}\right\|_2\left(\varepsilon_{\bf y}+\varepsilon_{\bf A}^{(K)}+\left(1+\varepsilon_{\bf y}\right)\left(r_K+s_K\right)\right).
\end{align}

Since
\begin{align}
\big\|{\bf x}-{\bf x}^{[l]}\big\|_2\le&\left\|{\bf \Psi}\right\|_2\big\|{\bf s}-{\bf s}^{[l]}\big\|_2,\label{proofequ41}\\
\left\|{\bf x}\right\|_2\ge&\left\|{\bf \Psi}^{-1}\right\|_2^{-1}\left\|{\bf s}\right\|_2\label{proofequ42},
\end{align}
together with (\ref{proofequ3}) and (\ref{proofequ7}), inequality (\ref{generalrelativeerror}) can be obtained.

For the second part of the theorem, noticing that when (\ref{generaliteration}) holds, it is obvious that
\begin{align}
a{\tilde C}^{l}+r_{K}\leq & \varepsilon_{\bf y}+\varepsilon_{\bf A}^{(K)}+r_{K}+s_{K}\nonumber\\
\leq & \varepsilon_{\bf y}+\varepsilon_{\bf A}^{(K)}+\left(1+\varepsilon_{\bf y}\right)\left(r_{K}+s_{K}\right)
\end{align}
and (\ref{generalrelativeerroriteration}) follows immediately, which completes the proof of Theorem~\ref{Greedyerrorth}.
\end{proof}

\subsection{Discussion}

In this subsection, several remarks are drawn to have a deep insight into the results of this paper.

{\bf Remark 1} As can be seen from (\ref{generalrelativeerroriteration}), the relative error of greedy pursuits with replacement is linear in both kinds of perturbations, and almost linear in the approximation error of $\bf s$ to a $K$-sparse vector. If $\bf s$ is $K$-sparse and the sensing process is non-perturbed, according to (\ref{sparserelativeerror}) with $\varepsilon_{\bf y}=\varepsilon_{\bf A}^{(K)}=0$, the relative error is bounded by an exponential decay function of iteration number, which indicates that the recovered solution can approach the original signal at any given precision. If $\bf s$ is $K$-sparse and $\bf x$ is sensed under general perturbations, (\ref{sparserelativeerroriteration}) demonstrates that the recovery accuracy is bounded by the size of both perturbations, and the recovery performance is stable in this scenario. If the sensing process is non-perturbed for general signal $\bf x$, according to (\ref{generalrelativeerroriteration}) with $\varepsilon_{\bf y}=\varepsilon_{\bf A}^{(K)}=0$, the recovery accuracy is determined by how well $\bf s$ can be approximated by a $K$-sparse vector.

{\bf Remark 2} Under the assumption (\ref{generalcondition}), the base $\tilde C$ of the exponential function in (\ref{generalrelativeerror}) is less than one, which guarantees the convergence of the recovery error. As ${\tilde \delta}_{bK}\rightarrow c$, the base $\tilde C$ approaches one and the constant $\tilde D$ approaches infinity. Also, due to the monotonicity of RICs, for less sparse signals with larger $K$, the constant $\tilde D$ gets larger.

For these three greedy pursuits with replacement, $K$ is quite a significant parameter and needs to be known a priori. An improper selection of $K$ may result in a great loss of recovery performance. If $K$ is selected too large, the condition (\ref{generalcondition}) can be hardly satisfied, and the parameter $\tilde D$ may be quite large. On the other hand, too small $K$ indicates large $r_K$ and $s_K$, which also leads to poor recovery performance.

As a typical compressible signal, strong-decaying signal $\bf s$ is a vector whose ordered coefficients in certain basis satisfy
\begin{align}
\left|{\bf s}\right|_{(l)}\geq p\left|{\bf s}\right|_{(l+1)},\ \ 1\leq l\leq N-1,
\end{align}
where $\left|{\bf s}\right|_{(l)}$ is the $l$-th largest magnitude among the elements of $\bf s$, and $p>1$ is a constant controls the speed of the decay: the larger $p$ is, the faster $\bf s$ decays. Simple calculation implies that
\begin{align}
r_K\leq p^{-K},\ \ s_K\leq \frac{\sqrt{p+1}}{\sqrt{K(p-1)}}p^{-K}.
\end{align}
For compressible signal satisfying a power law, i.e.,
\begin{align}\label{powerlaw}
\left|{\bf s}\right|_{(l)}\approx R\cdot l^{-p},\ \ 1\leq l\leq N,
\end{align}
where $R$ is a positive constant denoting the radius of weak-$\ell_{1/p}$ ball and $p>1$ controls the speed of the decay, it can also be calculated that
\begin{align}\label{compress}
r_K\approx K^{1/2-p},\ \ s_K\approx \frac{\sqrt{2p-1}}{p-1}K^{1/2-p}.
\end{align}

In practice, the parameter $K$ should be adjusted according to $\varepsilon_{\bf y}$, $\varepsilon_{\bf A}^{(K)}$, and the decay parameter $p$. In the scenario with relatively large perturbations, according to (\ref{generalrelativeerroriteration}), $K$ can be selected a small value such that ${\tilde\delta}_{bK}$ and $\tilde D$ are small, and $r_K$ and $s_K$ are comparable to $\varepsilon_{\bf y}$ and $\varepsilon_{\bf A}^{(K)}$. In the scenario with relatively small perturbations, the recovery error is mainly determined by $r_K$ and $s_K$, and $K$ is a trade-off parameter between these ratios and $\tilde D$.

{\bf Remark 3} Considering the sensing process (\ref{sysper1}) with ${\bf \tilde \Phi}={\bf \Phi}+{\bf \Delta}$ and ${\bf \tilde A}={\bf \tilde \Phi}{\bf \Psi}={\bf A+E}$,
\begin{align}
{\bf \tilde y}=&({\bf A}+{\bf E}){\bf s}+{\bf e}={\bf A}{\bf s}_K+\left({\bf e}+{\bf E}{\bf s}_K+{\bf \tilde A}{\bf s}_K^c\right).
\end{align}
Define ${\bf \tilde e}={\bf e}+{\bf E}{\bf s}_K+{\bf \tilde A}{\bf s}_K^c$ as the error term. Following the steps of proof of Theorem~\ref{Greedyerrorth}, it can be derived that if the matrix $\bf A$ satisfies RIP with $\delta_{bK}\le c$, then
\begin{align}
\big\|{\bf s}-{\bf s}^{[l]}\big\|_2\le a{C}^{l}\left\|{\bf s}\right\|_2+\left\|{\bf s}_K^c\right\|_2+{D}\left\|{\bf\tilde e}\right\|_2.
\end{align}

Since
\begin{align*}
\left\|{\bf \tilde e}\right\|_2\leq\left\|{\bf e}\right\|_2+\left\|{\bf E}\right\|_2^{(K)}\left\|{\bf s}\right\|_2+\big\|{\bf \tilde A}\big\|_2^{(K)}\left\|{\bf s}\right\|_2\left(r_K+s_K\right)
\end{align*}
and the fact that
\begin{align}
\big\|{\bf \tilde A}\big\|_2^{(K)}\le \|{\bf A}\|_2^{(K)}+\|{\bf E}\|_2^{(K)}\le \left(1+\varepsilon_{\bf A}^{(K)}\right)\|{\bf A}\|_2^{(K)},
\end{align}
it can be derived that the relative error of the solution in the $l$-th iteration obeys
\begin{align}
\frac{\displaystyle \big\|{\bf x}-{\bf x}^{[l]}\big\|_2}{\displaystyle \left\|{\bf x}\right\|_2}&\le\kappa_{\bf \Psi}\bigg(a{C}^{l}+r_{K}+{D}\sqrt{1+\delta_K}\left(\varepsilon_{\bf y}+\varepsilon_{\bf A}^{(K)}+\big(1+\varepsilon_{\bf y}+\varepsilon_{\bf A}^{(K)}\big)\left(r_{K}+s_{K}\right)\right)\bigg).
\end{align}

Notice that in this scenario, the relative error bound of greedy pursuits with replacement is also linear in the relative perturbations of both measurement vector and sensing matrix, which indicates that the recovery performance is stable against general perturbations. Also, it should be stressed that for both scenarios, the demands of RIP are all for the available matrix $\bf \tilde A$ or $\bf A$ in the process of recovery.

\section{Comparison with Oracle Recovery}

In this section, the upper and lower bounds of oracle recovery are derived, and are compared with those of greedy pursuits with replacement. It reveals that the results in this paper are optimal up to the coefficients.

\subsection{Error Bound of Oracle Recovery}

Consider the oracle recovery where the locations of the $K$ largest entries in magnitude of the vector $\bf s$ are known a priori. Assume $\rm S$ is the set of the $K$ locations, i.e., ${\rm S}={\rm supp}({\bf s}_K)$.

Recall the sensing process (\ref{sysper2}) where both perturbations exist. Through least squares (LS) method, the estimated solution ${\bf \hat x}={\bf \Psi}{\bf \hat s}$ of oracle recovery is obtained by \begin{align}
{\bf \hat s}_{\rm S}={\bf \tilde A}_{\rm S}^{\dagger}{\bf \tilde y},\quad {\bf \hat s}_{{\rm S}^c}={\bf 0}.
\end{align}
It is easy to check that
\begin{align}
{\bf \hat s}_{\rm S}={\bf \tilde A}_{\rm S}^{\dagger}{\bf \tilde y}={\bf \tilde A}_{\rm S}^{\dagger}\left({\bf \tilde A}{\bf s}_K+{\bf\tilde e}\right)={\bf s}_{\rm S}+{\bf \tilde A}_{\rm S}^{\dagger}{\bf\tilde e}.
\end{align}
Thus the estimation error of ${\bf \hat s}$ obeys
\begin{align}\label{lserrorequ}
\left\|{\bf s}-{\bf \hat s}\right\|_2^2&=\left\|{\bf s}_{\rm S}-{\bf \hat s}_{\rm S}\right\|_2^2+\left\|{\bf s}_{{\rm S}^c}-{\bf \hat s}_{{\rm S}^c}\right\|_2^2\nonumber\\
&=\big\|{\bf \tilde A}_{\rm S}^{\dagger}{\bf\tilde e}\big\|_2^2+\left\|{\bf s}_K^c\right\|_2^2.
\end{align}

According to (\ref{lserrorequ}), the estimation error obeys
\begin{align}\label{oracleequ}
\left\|{\bf s}-{\bf \hat s}\right\|_2&\leq\big\|{\bf \tilde A}_{\rm S}^{\dagger}{\bf\tilde e}\big\|_2+\left\|{\bf s}_K^c\right\|_2\nonumber\\
&\leq\big\|{\bf \tilde A}_{\rm S}^{\dagger}\big\|_2\left\|{\bf\tilde e}\right\|_2+\left\|{\bf s}_K^c\right\|_2.
\end{align}
Substituting (\ref{proofequ7}) into (\ref{oracleequ}), and together with (\ref{proofequ41}), (\ref{proofequ42}), and the fact that
\begin{align}
\big\|{\bf \tilde A}_{\rm S}^{\dagger}\big\|_2\leq\frac{1}{\sqrt{1-{\tilde\delta}_K}}\triangleq
{\hat D},
\end{align}
the relative error of LS solution is derived following the steps in the proof of Theorem~\ref{Greedyerrorth}
\begin{align}\label{LSrelativeerror}
\frac{\displaystyle \left\|{\bf x}-{\bf \hat x}\right\|_2}{\displaystyle \left\|{\bf x}\right\|_2}\leq&\kappa_{\bf \Psi}\left({\hat D}\sqrt{1+\delta_K}+1\right)\left(\varepsilon_{\bf y}+\varepsilon_{\bf A}^{(K)}+\left(1+\varepsilon_{\bf y}\right)\left(r_{K}+s_{K}\right)\right).
\end{align}

Comparing (\ref{LSrelativeerror}) with (\ref{generalrelativeerroriteration}), it can be derived that after finite iterations, the error bounds of greedy pursuits with replacement and the error bound of oracle recovery only differ in coefficients, and they are all of the same order on the noise level and approximation error
\begin{align}
O\left(\varepsilon_{\bf y}+\varepsilon_{\bf A}^{(K)}+\left(1+\varepsilon_{\bf y}\right)\left(r_{K}+s_{K}\right)\right)
\end{align}
under general perturbations.

\subsection{Lower Bound Analysis of Oracle Recovery}

According to (\ref{proofequ3}), (\ref{proofequ4}), and (\ref{proofequ41}), after
\begin{align*}
l=\left\lceil\log_{\tilde C}\left(\frac{\displaystyle \|{\bf e}\|_2+\|{\bf E}\|_2^{(K)}\|{\bf s}_K\|_2}{\displaystyle a\|{\bf A}\|_2^{(K)}\|{\bf s}\|_2}\right)\right\rceil
\end{align*}
iterations, the recovery error of greedy pursuits with replacement satisfies
\begin{align}\label{oracleequ6}
&\big\|{\bf x}-{\bf x}^{[l]}\big\|_2\le\left\|{\bf \Psi}\right\|_2\left({\tilde D}+\frac{1}{\sqrt{1-\delta_K}}\right)\left(\left\|{\bf e}\right\|_2+\left\|{\bf E}\right\|_2^{(K)}\left\|{\bf s}_K\right\|_2+\left\|{\bf A}\right\|_2^{(K)}\left\|{\bf s}\right\|_2\left(r_K+s_K\right)\right).
\end{align}
To show the optimality of the results in this paper, we prove that for some certain $\bf s$, $\bf e$, and $\bf E$, the lower bound of the estimation error of oracle recovery is also linear in the following three terms
\begin{align}\label{errorterms}
\left\|{\bf e}\right\|_2,\quad\left\|{\bf E}\right\|_2^{(K)}\left\|{\bf s}_K\right\|_2,\quad\left\|{\bf s}\right\|_2\left(r_K+s_K\right),
\end{align}
which quantify measurement noise, system perturbation, and approximation error, respectively.

According to (\ref{lserrorequ}), the estimation error of ${\bf \hat s}$ obeys
\begin{align}\label{oracleequ1}
\left\|{\bf s}-{\bf \hat s}\right\|_2\geq\frac{1}{\sqrt{2}}\left(\big\|{\bf \tilde A}_{\rm S}^{\dagger}{\bf\tilde e}\big\|_2+\left\|{\bf s}_K^c\right\|_2\right).
\end{align}
The system perturbation $\bf E$ can be chosen as ${\bf E}=\varepsilon_{\bf A}^{(K)}${\bf A}, and the vector ${\bf s}_K$ is selected such that
\begin{align}\label{oracleequ2}
\left\|{\bf E}{\bf s}_K\right\|_2=\left\|{\bf E}\right\|_2^{(K)}\left\|{\bf s}_K\right\|_2.
\end{align}
Meanwhile, the vector ${\bf s}_K^c$ is also assumed $K$-sparse, and it can be assumed that
\begin{align}\label{oracleequ7}
\left(-{\bf E}{\bf s}_K\right)^{\rm T}{{\bf A}_{\rm S}}{\bf A}_{\rm S}^{\rm T}\left({\bf A}{\bf s}_K^c\right)\geq0,
\end{align}
otherwise $\left(-{\bf s}_K^c\right)$ is applied. Furthermore, the measurement noise $\bf e$ is chosen in the column space of ${\bf A}_{\rm S}$, and satisfies
\begin{align}\label{oracleequ8}
{\bf e}^{\rm T}{{\bf A}_{\rm S}}{\bf A}_{\rm S}^{\rm T}\left(-{\bf E}{\bf s}_K+{\bf A}{\bf s}_K^c\right)\ge0.
\end{align}

Due to the above assumptions and the fact that
\begin{align*}
{\bf \tilde A}={\bf A+E}=\left(1+\varepsilon_{\bf A}^{(K)}\right){\bf A},
\end{align*}
it can be derived that
\begin{align}\label{oracleequ3}
\big\|{\bf \tilde A}_{\rm S}^{\dagger}{\bf\tilde e}\big\|_2=\big\|\big({\bf \tilde A}_{\rm S}^{\rm T}{\bf \tilde A}_{\rm S}\big)^{-1}{\bf \tilde A}_{\rm S}^{\rm T}{\bf\tilde e}\big\|_2\ge\frac{1}{1+\tilde \delta_K}\big\|{\bf A}_{\rm S}^{\rm T}{\bf\tilde e}\big\|_2.
\end{align}
According to the definition of ${\bf\tilde e}$, (\ref{oracleequ7}), and (\ref{oracleequ8}), it can be calculated that
\begin{align}\label{oracleequ9}
\big\|{\bf A}_{\rm S}^{\rm T}{\bf\tilde e}\big\|_2&\ge\frac{1}{\sqrt{2}}\left(\left\|{\bf A}_{\rm S}^{\rm T}{\bf e}\right\|_2+\left\|{\bf A}_{\rm S}^{\rm T}\big(-{\bf E}{\bf s}_K+{\bf A}{\bf s}_K^c\big)\right\|_2\right)\nonumber\\
&\ge\frac{1}{\sqrt{2}}\left(\left\|{\bf A}_{\rm S}^{\rm T}{\bf e}\right\|_2+\left\|{\bf A}_{\rm S}^{\rm T}{\bf E}{\bf s}_K\right\|_2\right).
\end{align}
Since $\bf e$ and ${\bf E}{\bf s}_K$ both belong to the column space of ${\bf A}_{\rm S}$, (\ref{oracleequ9}) and (\ref{oracleequ2}) further imply
\begin{align}\label{oracleequ10}
\big\|{\bf A}_{\rm S}^{\rm T}{\bf\tilde e}\big\|_2&\ge\frac{\sqrt{1-\delta_K}}{\sqrt{2}}\left(\left\|{\bf e}\right\|_2+\left\|{\bf E}{\bf s}_K\right\|_2\right)\nonumber\\
&=\frac{\sqrt{1-\delta_K}}{\sqrt{2}}\left(\left\|{\bf e}\right\|_2+\left\|{\bf E}\right\|_2^{(K)}\left\|{\bf s}_K\right\|_2\right).
\end{align}
In addition, Cauchy-Schwartz inequality implies
\begin{align}\label{oracleequ11}
\left\|{\bf s}_K^c\right\|_2\geq\frac{1}{2}\left(\left\|{\bf s}_K^c\right\|_2+\frac{\displaystyle \left\|{\bf s}_K^c\right\|_1}{\displaystyle \sqrt{K}}\right)=\frac{1}{2}\left\|{\bf s}\right\|_2\left(r_K+s_K\right).
\end{align}

According to (\ref{oracleequ1}), (\ref{oracleequ3}), (\ref{oracleequ10}), (\ref{oracleequ11}), and the fact that
\begin{align*}
\big\|{\bf x}-{\bf\hat x}\big\|_2\ge&\left\|{\bf \Psi}^{-1}\right\|_2^{-1}\big\|{\bf s}-{\bf\hat s}\big\|_2,
\end{align*}
the lower bound of the estimation error of oracle recovery satisfies
\begin{align}
\big\|{\bf x}-{\bf\hat x}\big\|_2\ge\frac{\left\|{\bf \Psi}^{-1}\right\|_2^{-1}}{2}\Bigg(&\frac{\sqrt{1-\delta_K}}{1+\tilde\delta_K}\left(\left\|{\bf e}\right\|_2+\left\|{\bf E}\right\|_2^{(K)}\left\|{\bf s}_K\right\|_2\right)+\frac{1}{\sqrt{2}}\left\|{\bf s}\right\|_2\left(r_K+s_K\right)\Bigg),
\end{align}
which is also linear in the three terms (\ref{errorterms}).

According to the above analysis, it can be concluded that there exist ${\bf s}$, $\bf e$, and $\bf E$ with which the results in this paper are essentially tight up to the coefficients, i.e., the performances of these algorithms and oracle recovery are of the same order. In general, no recovery algorithms can do better than the oracle least squares method. Thus, the greedy pursuits with replacement, CoSaMP, SP, and IHT, can provide oracle-order recovery performance against general perturbations.

\section{Numerical Simulations}

Three numerical simulations are performed in MATLAB and demonstrated as follows. The first simulation compares the relative error of greedy pursuits with replacement and oracle LS method for different sparsity levels. The second simulation compares the relative error with different relative perturbations. The final simulation considers compressible signals and presents the influence of parameter $K$ on the relative recovery error.

In each trial the matrix $\bf \Phi$ of size $512\times2048$ is randomly generated with independent Gaussian distributed entries with variance $\sigma^2=1/512$ so that the expectation of the $\ell_2$ norm of each column is normalized. The basis matrix $\bf \Psi$ is the identity matrix, i.e., ${\bf x=s}$ and ${\bf A=\Phi}$. The nonzero locations of vector $\bf s$ are randomly selected among all possible choices, and the nonzero entries are generated independently from normal distribution. Then, for each pair of relative perturbations, ${\bf \tilde y}$ and ${\bf \tilde \Phi}$ are generated according to (\ref{sysper2}) and (\ref{errorboundsss}). Notice that only $\varepsilon_{\bf A}$ is used in the simulations since calculating $\varepsilon_{\bf A}^{(K)}$ is NP-hard. However, it is reasonable because $\varepsilon_{\bf A}\approx \varepsilon_{\bf A}^{(K)}$ holds for all $K$ with high probability when both $\bf A$ and $\bf E$ are random Gaussian matrices \cite{Matthew}.

\begin{figure}[t]
\begin{center}
\includegraphics[width=4in]{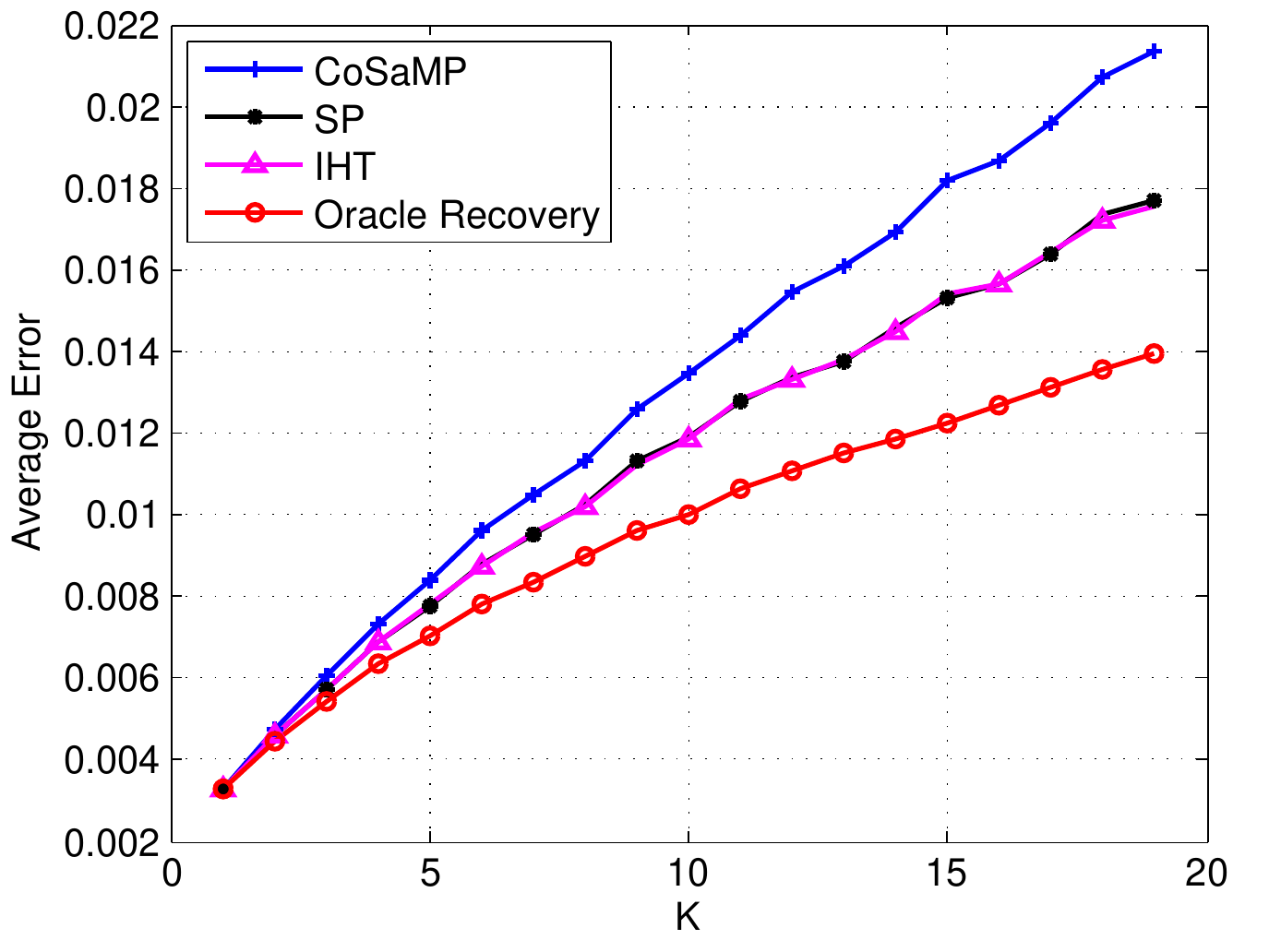}
\caption{Average relative error (1000 trials) of greedy pursuits with replacement and oracle LS recovery versus sparsity level $K$ when $\varepsilon_{\bf A}=0.05$ and $\varepsilon_{\bf y}=0.05$.}\label{different_K}
\end{center}
\end{figure}

\begin{figure*}[tp]
\begin{center}
\includegraphics[width=\textwidth]{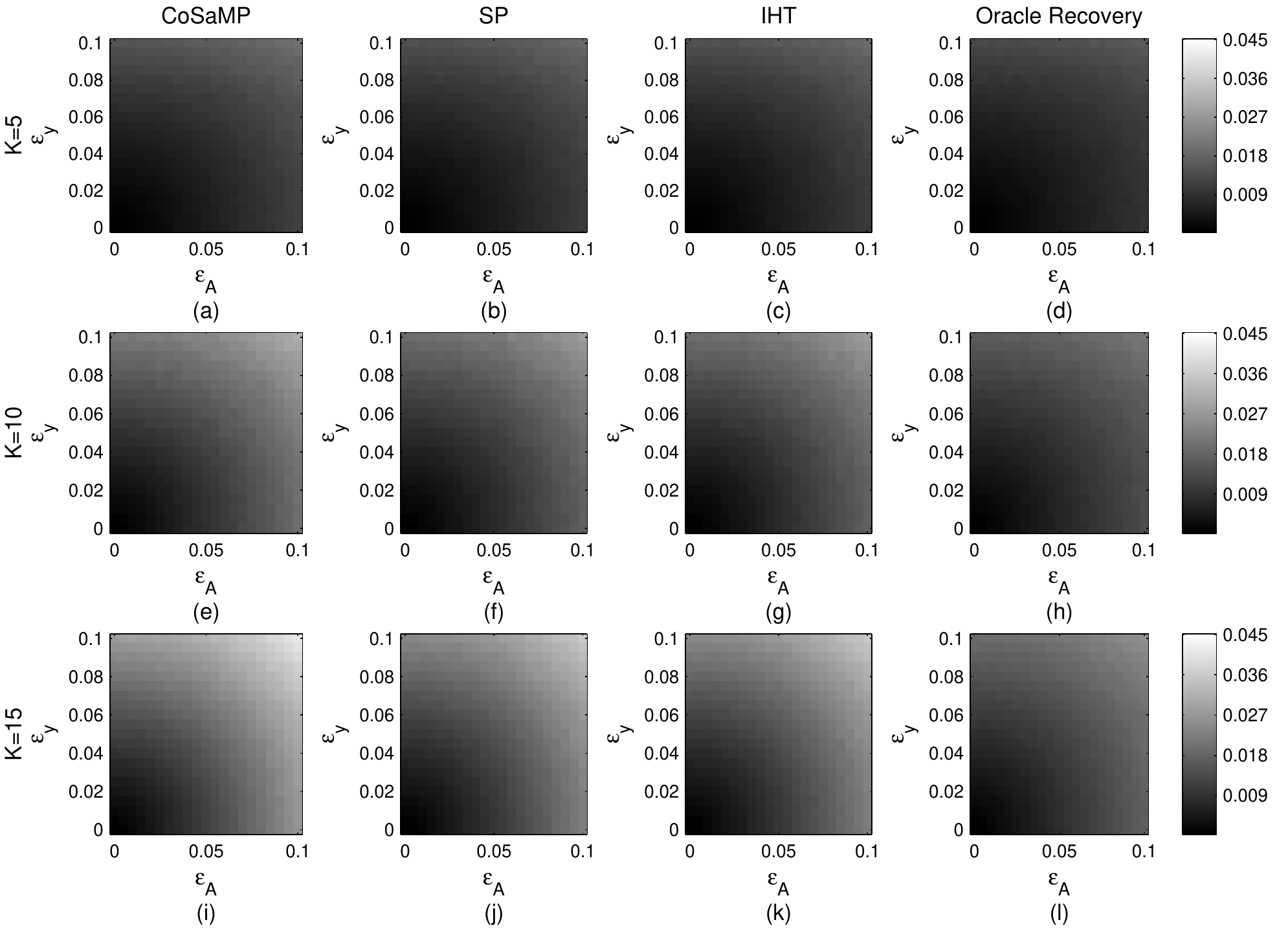}
\caption{Average relative error (1000 trials) of greedy pursuits with replacement and oracle LS recovery versus different $\varepsilon_{\bf A}$ and $\varepsilon_{\bf y}$. The results are represented for the same $K$ in the same row, and for the same recovery algorithm in the same column, with labels in the left and on the top, respectively.}\label{differenterror}
\end{center}
\end{figure*}

In the first experiment, the relative perturbations are set to $\varepsilon_{\bf A}=0.05$ and $\varepsilon_{\bf y}=0.05$. The sparsity level $K$ varies from $1$ to $20$, and the simulation is repeated for $1000$ times. As can be seen from Fig.~\ref{different_K}, when $K< 20$, the relative error increases almost linearly as $K$ increases. When the sparsity level $K$ continues to increase, the relative error of IHT cannot be bounded any more, which indicates that the RIP condition is violated.

\begin{figure}[t]
\begin{center}
\includegraphics[width=4in]{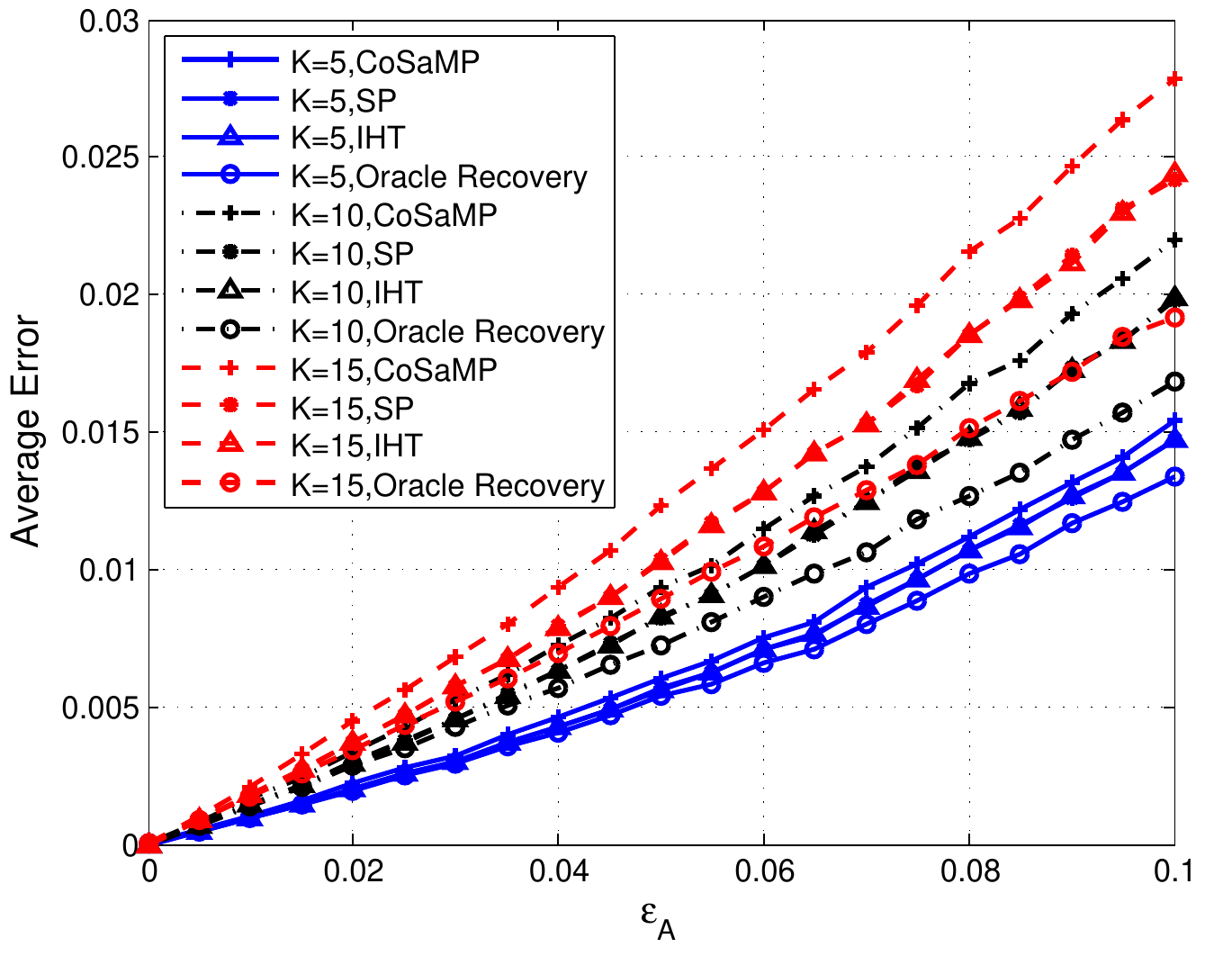}
\caption{Average relative error (1000 trials) of greedy pursuits with replacement and oracle LS recovery versus $\varepsilon_{\bf A}$ for different sparsity level $K$ when $\varepsilon_{\bf y}=0$.}\label{systemerror}
\end{center}
\end{figure}

In the second experiment, for fixed sparsity level $K=5$, $10$, and $15$, the relative perturbations $\varepsilon_{\bf A}$ and $\varepsilon_{\bf y}$ both vary from $0$ to $0.1$ with step size 0.005. The simulation is conducted 1000 trials to obtain the average relative error. The results are demonstrated in Fig.~\ref{differenterror}, where they are represented for the same $K$ in the same row, and for the same recovery algorithm in the same column, with labels in the left and on the top, respectively. As can be seen from them, for the same $K$, the surface of oracle LS method lies below the others, while the surface of CoSaMP is above on the top. The surfaces of SP and IHT are almost the same in the middle. It verifies that the relative errors of CoSaMP, SP, and IHT are linear in the measurement perturbation as well as system perturbation, and these greedy pursuits with replacement can provide oracle-order recovery performance. It needs to be pointed out that for the same recovery algorithm, the gradients of these surfaces increase as the increase of $K$, which confirms that the coefficients of these error bounds are related to $K$ as discussed in Section III and IV. To display the results more intuitively, the average relative error is plotted versus $\varepsilon_{\bf A}$ for different $K$ when $\varepsilon_{\bf y}=0$, as in Fig.~\ref{systemerror}. As can be seen, the relative error scales almost linearly as $\varepsilon_{\bf A}$, and the slopes of these curves increase as $K$ for the same recovery algorithm.

\begin{figure}[t]
\begin{center}
\includegraphics[width=4in]{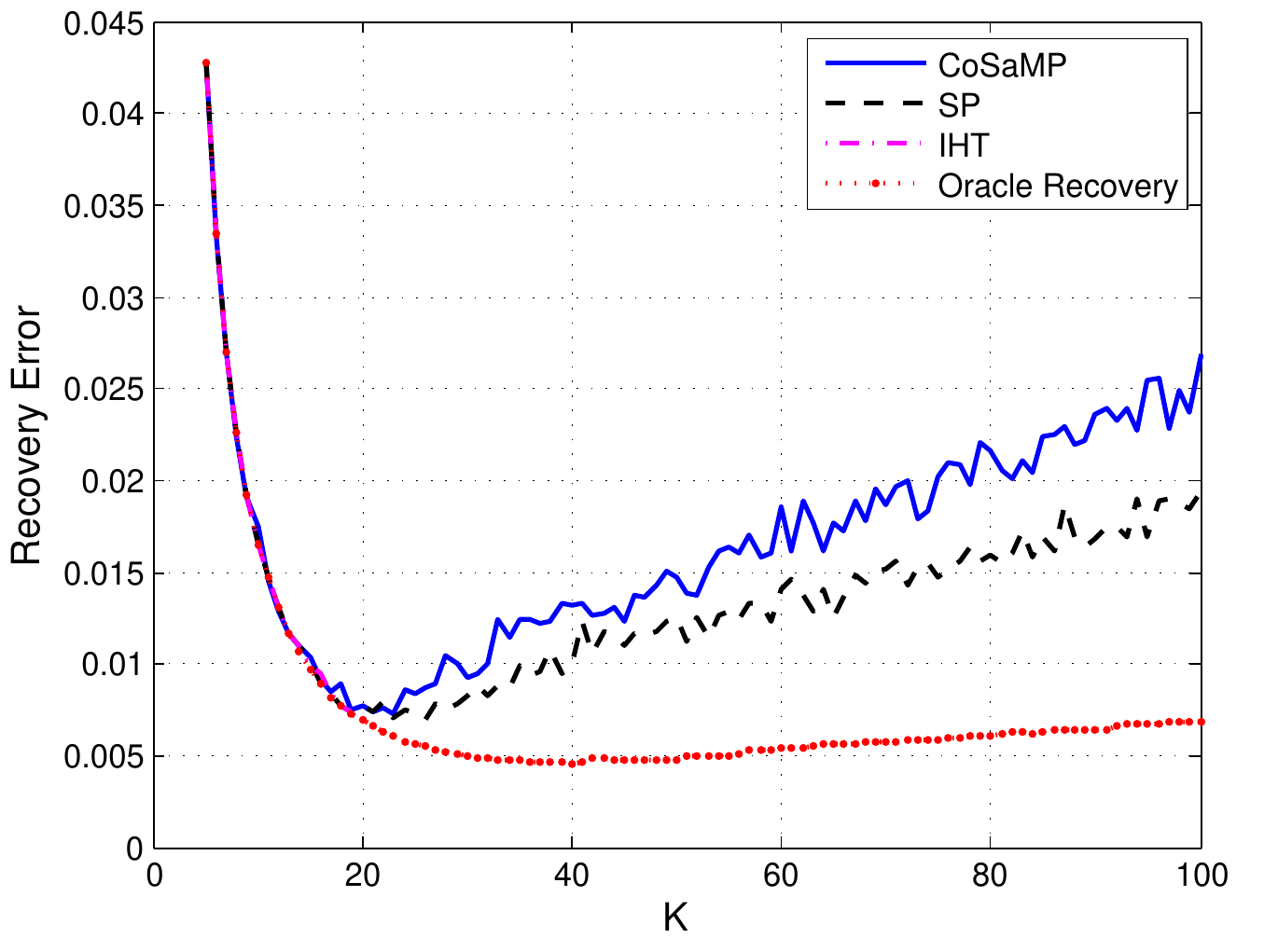}
\caption{Relative error of greedy pursuits with replacement and oracle LS recovery for compressible signal versus different parameter $K$ when $\varepsilon_{\bf A}=0.01$ and $\varepsilon_{\bf y}=0.01$. IHT algorithm becomes invalid when $K\ge20$, thus it is not plotted for the wider range.}\label{compressible}
\end{center}
\end{figure}

In the final experiment, we consider compressible signal satisfying power law (\ref{powerlaw}) with $R=1$ and $p=2$. The relative perturbations are set to $\varepsilon_{\bf A}=0.01$ and $\varepsilon_{\bf y}=0.01$. The parameter $K$ in the recovery algorithms varies from 5 to 100, and the results are plotted in Fig.~\ref{compressible}. Since IHT fails when $K\ge 20$, the recovery error of it is not shown for $K\ge 20$ in the figure. As can be seen from Fig.~\ref{compressible}, the optimal $K$ is around 22 for CoSaMP and SP, 19 for IHT, and around 40 for oracle recovery. Too small $K$ or large $K$ will affect the recovery performance for compressible signals, as is discussed in Remark~2. According to (\ref{compress}), for $K=22$ and $p=2$, the approximation error $r_K$ is about 0.010 and $s_K$ is about 0.017, which are comparable to $\varepsilon_{\bf A}$ and $\varepsilon_{\bf y}$. This further confirms the analysis in Remark~2 that the parameter $K$ should be adjusted in accordance with $\varepsilon_{\bf A}^{(K)}$ and $\varepsilon_{\bf y}$.

\section{Related Works}

A similar concept of this paper is presented in \cite{Giryes}, where the authors also establish a oracle-order performance guarantee for CoSaMP, SP, and IHT algorithms. Based on RIP, the analysis considers the recovery of a $K$-sparse signal with the assumption that the measurement vector is corrupted by additive random white Gaussian noise. The main result of \cite{Giryes} is stated as follows, where some notations are replaced for the sake of consistency to our work.

Assume that the white Gaussian noise vector is with covariance matrix $\sigma^2{\bf I}$, and that the columns of matrix $\bf A$ are normalized. Under certain conditions of RICs and with probability exceeding $1-(\sqrt{\pi(1+a)\log N}\cdot N^a)^{-1}$, it holds that
\begin{align}\label{relatedoracle}
\|{\bf x}-{\bf \hat x}\|_2^2\le 2C^2(1+a)\log N\cdot K\sigma^2,
\end{align}
where ${\bf \hat x}$ is the recovered signal, and $C$ is a constant related to specific recovery algorithms, sensing matrix, and sparsity level. The result is similar to those for the dantzig-selector and the basis pursuit, but with different constants. The $\log N$ factor in (\ref{relatedoracle}) is proven to be unavoidable in \cite{logoptimal}, therefore this bound is optimal up to a constant factor. The result is also extended to the nearly-sparse case in \cite{Giryes}.

In other relevant works \cite{Herman,Wang}, the error bounds of CoSaMP and SP are also derived under general perturbations. The analysis is performed following the steps of \cite{Matthew} for basis pursuit. Define
\begin{align*}
\alpha_K= \frac{\|{\bf x}-{\bf x}_K\|_2}{\|{\bf x}_K\|_2},\ \ \beta_K= \frac{\|{\bf x}-{\bf x}_K\|_1}{\sqrt{K}\|{\bf x}_K\|_2},\ \ \kappa_{\bf A}^{(K)}=
\frac{\sqrt{1+\delta_K}}{\sqrt{1-\delta_K}}.
\end{align*}
The main result in \cite{Herman} states that under the conditions that
\begin{align}\label{CoSaMPRIP}
\delta_{4K}\le \frac{1.1}{(1+\varepsilon_{\bf A}^{(4K)})^2}-1
\end{align}
and
\begin{align}\label{CoSaMPkappa}
\alpha_K+\beta_K\le \frac{1}{2\kappa_{\bf A}^{(K)}},
\end{align}
the recovered solution of CoSaMP satisfies
\begin{align}
\|{\bf x}-{\bf \hat x}\|_2\le C\cdot\Big(&\|{\bf x}-{\bf x}_K\|_2+\frac{1}{\sqrt{K}}\|{\bf x}-{\bf x}_K\|_1+\|{\bf e}\|_2+(\|{\bf E}\|_2\alpha_K+\|{\bf E}\|_2^{(K)})\|{\bf y}\|_2\Big).
\end{align}
Further define
\begin{align}
\varepsilon_{{\bf A},K,{\bf y}}\triangleq \left(\frac{\varepsilon_{\bf A}^{(K)}\kappa_{\bf A}^{(K)}+\varepsilon_{\bf A}\gamma_{\bf A}\alpha_K}{1-\kappa_{\bf A}^{(K)}(\alpha_K+\beta_K)}+\varepsilon_{\bf y}\right)\|{\bf y}\|_2,
\end{align}
where
\begin{align}
\gamma_{\bf A}=\frac{\|{\bf A}\|_2}{\sqrt{1-\delta_K}}.
\end{align}
The main result in \cite{Wang} states that under the conditions that
\begin{align}\label{SPRIP}
\delta_{3K}\le \frac{1.083}{(1+\varepsilon_{\bf A}^{(3K)})^2}-1
\end{align}
and that
\begin{align}\label{SPkappa}
\alpha_K+\beta_K< \frac{1}{\kappa_{\bf A}^{(K)}},
\end{align}
the recovered solution of SP satisfies
\begin{align}
\|{\bf x}-{\bf \hat x}\|_2\le& (C''_K+1)\|{\bf x}-{\bf x}_K\|_2+C''_K\frac{\|{\bf x}-{\bf x}_K\|_1}{\sqrt{K}}+C'_K\varepsilon_{{\bf A},K,{\bf y}}.
\end{align}

Our analysis, on the other hand, derives a unified form of the relative error bounds of CoSaMP, SP, and IHT under general perturbations. Specified in Theorem~\ref{Greedyerrorth}, for compressible signals, when measurement vector and sensing matrix are perturbed, under the condition of (\ref{generalcondition}), the relative error bound of recovered solution in each iteration is derived as (\ref{generalrelativeerror}). It is also proved that after finite iterations, the error bound is linear in both perturbations and almost linear in the approximation error, as in (\ref{generalrelativeerroriteration}). The result is compared with oracle recovery, where the locations of $K$ largest entries in magnitude are known a priori. It is also proved that for some certain $\bf x$, $\bf e$, and $\bf E$, the lower bound of oracle recovery only differs in coefficients from the error bounds of greedy pursuits with replacement. Therefore, the oracle-order recovery performance of them is guaranteed in this work.

The main difference between \cite{Giryes} and our work is that we consider a more general completely perturbed scenario, and the optimality of the recovery performance is also in this sense. Compared with \cite{Herman,Wang}, the demand of RICs in our work is for the perturbed matrix $\bf \tilde A$ other than $\bf A$, which is due to the fact that only $\bf \tilde A$ is available for recovery. Also, the demand of RICs is with a constant parameter here. In addition, the condition such as (\ref{CoSaMPkappa}) or (\ref{SPkappa}) is not required in our assumption. Our results are compared with oracle recovery and shown to be optimal up to the coefficients, and they are verified by plentiful numerical simulations.

\section{Conclusion}

In this paper, the relative error bounds of three greedy pursuits with replacement under general perturbations are derived. It is shown that these bounds are linear in both perturbations, which leads to the stability of these algorithms against general perturbations. Furthermore, these error bounds are compared with that of oracle LS solution with the locations of $K$ largest entries in magnitude known a priori. Since the error bounds of CoSaMP, SP, and IHT algorithms are of the same order as the lower bound of oracle solution for some certain signal and perturbations, it can be concluded that these greedy pursuits with replacement can provide oracle-order recovery performance against general perturbations. Numerical simulations verify that the relative recovery errors of these algorithms are linear in both perturbations, and they exhibit oracle-order recovery performance. Discussions and simulations also reveal how moderate parameter $K$ achieves better recovery performance for compressible signals.

\appendix

\appendixtitleon
\begin{appendices}

\section{Discussions of Algorithms}

Greedy pursuits with replacement are briefly introduced in Section~II-B. The pseudo codes of CoSaMP, SP, and IHT are demonstrated in Table~\ref{CoSaMPcode}, Table~\ref{SPcode}, and Table~\ref{IHTcode}, respectively. The definitions of the symbols are presented in Section~III-A, and ${\rm supp}({\bf u})$ denotes the support set of $\bf u$.

\begin{table}[t]
\renewcommand{\arraystretch}{1.2}
\caption{The Pseudo Code of CoSaMP}
\begin{center}
\begin{tabular}{l}
\toprule[1pt] {\bf Initialization}\\\label{CoSaMPcode}
\hspace{1.5em}1. $l=0$;\\
\hspace{1.5em}2. ${\bf s}^{[0]}={\bf 0}$, ${\bf r}={\bf y}$;\\
\hline{\bf Repeat}\\
\hspace{1.5em}3. $l=l+1$;\\
\hspace{1.5em}4. ${\bf u}={\bf A}^{\rm T}{\bf r}$, $\Omega={\rm supp}\left({\bf u}_{2K}\right)$;\\
\hspace{1.5em}5. ${\rm\tilde S}=\Omega\cup {\rm supp}\left({\bf s}^{[l-1]}\right)$;\\
\hspace{1.5em}6. ${\bf b}_{\rm\tilde S}={\bf A}_{\rm\tilde S}^{\dagger}{\bf y}$, ${\bf b}_{{\rm\tilde S}^c}={\bf 0}$;\\
\hspace{1.5em}7. ${\bf s}^{[l]}={\bf b}_K$, ${\bf r}={\bf y}-{\bf A}{\bf s}^{[l]}$;\\
{\bf Until} stopping criterion is satisfied\\
\bottomrule[1pt]
\end{tabular}
\end{center}
\end{table}

\begin{table}[t]
\renewcommand{\arraystretch}{1.2}
\caption{The Pseudo Code of SP}
\begin{center}
\begin{tabular}{l}
\toprule[1pt] {\bf Initialization}\\\label{SPcode}
\hspace{1.5em}1. $l=0$;\\
\hspace{1.5em}2. ${\bf u}={\bf A}^{\rm T}{\bf y}$, ${\rm S}^0={\rm supp}\left({\bf u}_{K}\right)$;\\
\hspace{1.5em}3. ${\bf s}^{[0]}_{{{\rm S}^0}}={\bf A}_{{\rm S}^0}^{\dagger}{\bf y}$, ${\bf s}^{[0]}_{({\rm S}^0)^c}={\bf 0}$;\\
\hspace{1.5em}4. ${\bf y}_r^0={\bf y}-{\bf A}{\bf s}^{[0]}$;\\
\hline{\bf Repeat}\\
\hspace{1.5em}5. $l=l+1$;\\
\hspace{1.5em}6. ${\bf u}={\bf A}^{\rm T}{\bf y}_r^{l-1}$, $\Omega={\rm supp}\left({\bf u}_{K}\right)$;\\
\hspace{1.5em}7. ${\rm \tilde S}^l=\Omega\cup {\rm S}^{l-1}$;\\
\hspace{1.5em}8. ${\bf s}_p={\bf A}_{{\rm \tilde S}^l}^{\dagger}{\bf y}$, ${\rm S}^l={\rm supp}\left(\left({\bf s}_p\right)_{K}\right)$;\\
\hspace{1.5em}9. ${\bf s}^{[l]}_{{{\rm S}^l}}={\bf A}_{{\rm S}^l}^{\dagger}{\bf y}$, ${\bf s}^{[l]}_{({\rm S}^l)^c}={\bf 0}$;\\
\hspace{1.5em}10. ${\bf y}_r^l={\bf y}-{\bf A}{\bf s}^{[l]}$;\\
{\bf Until} stopping criterion is satisfied\\
\bottomrule[1pt]
\end{tabular}
\end{center}
\end{table}

\begin{table}[t]
\renewcommand{\arraystretch}{1.2}
\caption{The Pseudo Code of IHT}
\begin{center}
\begin{tabular}{l}
\toprule[1pt] {\bf Initialization}\\\label{IHTcode}
\hspace{1.5em}1. $l=0$;\\
\hspace{1.5em}2. ${\bf s}^{[0]}={\bf 0}$;\\
\hline{\bf Repeat}\\
\hspace{1.5em}3. $l=l+1$;\\
\hspace{1.5em}4. ${\bf u}={\bf A}^{\rm T}\left({\bf y}-{\bf A}{\bf s}^{[l-1]}\right)$;\\
\hspace{1.5em}5. ${\bf s}^{[l]}=\left({\bf s}^{[l-1]}+{\bf u}\right)_K$;\\
{\bf Until} stopping criterion is satisfied\\
\bottomrule[1pt]
\end{tabular}
\end{center}
\end{table}

Based on Theorem~\ref{GPRa}, two remarks are derived as follows.

{\bf Remark 4} In the noiseless scenario, after finite iterations, the recovered solutions of CoSaMP and SP are guaranteed to be identical to the sparse signal ${\bf s}$. The result can be verified through the following statement.

Suppose $s_{\min}$ is the smallest magnitude of the nonzero entries of ${\bf s}$. Then after
\begin{align*}
l=\left\lfloor \log_{C} \left(\frac{\displaystyle s_{\min}}{\displaystyle a\left\|{\bf s}\right\|_2}\right)\right\rfloor+1
\end{align*}
iterations, the recovered solution ${\bf s}^{[l]}$ obeys
\begin{align}\label{Remarkequ1}
\big\|{\bf s}-{\bf s}^{[l]}\big\|_2< s_{\min}.
\end{align}
If the support of $\bf s$ is perfectly recovered, then for CoSaMP and SP, the solution is already identical to $\bf s$. Otherwise, at least one nonzero entry is not detected, thus the recovery error is no less than $s_{\min}$, which contradicts (\ref{Remarkequ1}). Notice that the solution of IHT does not possess the above property, since exact support recovery for IHT does not imply exact signal recovery.

{\bf Remark 5} According to (\ref{GPRerror2}), after
\begin{align*}
l=\left\lceil \log_C \left(\frac{\displaystyle \left\|{\bf e}\right\|_2}{\displaystyle a\left\|{\bf s}\right\|_2}\right)\right\rceil
\end{align*}
iterations, the error bound of the recovered solution satisfies
\begin{align}
\big\|{\bf s}-{\bf s}^{[l]}\big\|_2\le (D+1)\left\|{\bf e}\right\|_2,
\end{align}
which means that the error bound is proportional to the $\ell_2$ norm of the noise, and the recovery performance of greedy pursuits with replacement is stable in this scenario.

\section{Proof of Theorem~\ref{GPRa}}

\begin{proof}
For CoSaMP algorithm, the inequality (\ref{GPRerror1}) can be obtained by following the steps of the proof of Theorem~4.1 in \cite{CoSaMP}, while preserving RICs during the derivation. As for the second part of the theorem, it is easy to derive that if $\delta_{4K}<0.171$, then $C<1$. By recursion, it can be proved from (\ref{GPRerror1}) that
\begin{align}
\big\|{\bf s}-{\bf s}^{[l]}\big\|_2\le C^l\big\|{\bf s}-{\bf s}^{[0]}\big\|_2 +\frac{C_1}{1-C}\left\|{\bf e}\right\|_2,
\end{align}
which arrives (\ref{GPRerror2}).

For IHT algorithm, the inequality (\ref{GPRerror1}) is directly derived in the proof of Theorem~5 in \cite{IHT}. Similar to CoSaMP, the second part of the theorem can also be proved from (\ref{GPRerror1}).

For SP algorithm, the result cannot be directly found in \cite{SP}. However, it can be proved based on several facts about SP in \cite{SP}, and they are stated here without proofs.

\begin{lemma}\label{SPlemma1}
(Lemma~3 in \cite{SP}) Let ${\bf s}\in\mathbb{R}^N$ be a $K$-sparse vector, and ${\bf\tilde y}={\bf As}+{\bf e}$ is a noisy measurement vector where ${\bf A}\in\mathbb{R}^{M\times N}$ satisfies the RIP with parameter $\delta_{3K}$. For an arbitrary set ${\rm \hat S}\subset\{1,\ldots,N\}$ such that $|{\rm \hat S}|\leq K$, define ${\bf \hat s}$ as
\begin{align}
{\bf\hat s}_{\rm \hat S}={\bf A}_{\rm \hat S}^{\dagger}{\bf\tilde y},\ \ {\bf\hat s}_{{\rm \hat S}^c}={\bf 0}.
\end{align}
Then
\begin{align}
\left\|{\bf s}-{\bf \hat s}\right\|_2\leq\frac{\displaystyle 1}{\displaystyle 1-\delta_{3K}}\left\|{\bf s}_{{\rm S}-{\rm \hat S}}\right\|_2+\frac{\displaystyle 1+\delta_{3K}}{\displaystyle 1-\delta_{3K}}\left\|{\bf e}\right\|_2.
\end{align}
\end{lemma}

\begin{lemma}\label{SPlemma2}
(Theorem~10 in \cite{SP}) It holds that
\begin{align}
\left\|{\bf s}_{{\rm S}-{\rm \tilde S}^l}\right\|_2&\leq\frac{\displaystyle 2\delta_{3K}}{\displaystyle \left(1-\delta_{3K}\right)^2}\left\|{\bf s}_{{\rm S} -{\rm S}^{l-1}}\right\|_2+\frac{\displaystyle 2\left(1+\delta_{3K}\right)}{\displaystyle 1-\delta_{3K}}\left\|{\bf e}\right\|_2,\nonumber\\
\left\|{\bf s}_{{\rm S}-{\rm S}^l}\right\|_2&\leq \frac{\displaystyle 1+\delta_{3K}}{\displaystyle 1-\delta_{3K}}\left\|{\bf s}_{{\rm S}-{\rm \tilde S}^l}\right\|_2 +\frac{\displaystyle 2}{\displaystyle 1-\delta_{3K}}\left\|{\bf e}\right\|_2.\nonumber
\end{align}
Therefore
\begin{align}
\left\|{\bf s}_{{\rm S}-{\rm S}^l}\right\|_2\leq \frac{\displaystyle 2\delta_{3K}+2\delta_{3K}^2}{\displaystyle (1-\delta_{3K})^3}\left\|{\bf s}_{{\rm S}-{\rm S}^{l-1}}\right\|_2+\frac{\displaystyle 4(1+\delta_{3K})}{\displaystyle (1-\delta_{3K})^2}\left\|{\bf e}\right\|_2.\nonumber
\end{align}
\end{lemma}

Lemma~\ref{SPlemma2} directly implies inequality (\ref{SPerror1}). It's easy to check that if $\delta_{3K}\le 0.206$, then $C<1$, and
\begin{align}
\left\|{\bf s}_{{\rm S}-{\rm S}^l}\right\|_2&\le C^l\left\|{\bf s}_{{\rm S}-{\rm S}^{0}}\right\|_2+\frac{\displaystyle C_1}{\displaystyle 1-C}\left\|{\bf e}\right\|_2\nonumber\\
&\leq C^l\left\|{\bf s}\right\|_2+\frac{\displaystyle C_1}{\displaystyle 1-C}\left\|{\bf e}\right\|_2.\label{SPapp1}
\end{align}
Applying Lemma~\ref{SPlemma1} to (\ref{SPapp1}) and with the fact that $\delta_{3K}\le 0.206$, inequality (\ref{GPRerror2}) can be derived.
\end{proof}

\end{appendices}

\end{document}